\documentclass{PoS}

\PoS{PoS(LAT2005)178}

\usepackage{natbib,graphicx}
\bibpunct{[}{]}{,}{n}{}{;}

\title{Grand Canonical Potential for a Static Quark-Antiquark Pair at finite chemical potential 
}

\ShortTitle{Grand Canonical Potential for a Static $q\bar{q}$ Pair at $\mu\neq0$ }

\author{Zoltan Fodor\\
Department of Physics, University of Wuppertal, Gauss 20, D-42119, Germany \&  \\
Institute for Theoretical Physics, E$\ddot{\mbox{o}}$tv$\ddot{\mbox{o}}$s University, P$\acute{\mbox{a}}$zm$\acute{\mbox{a}}$ny 1, H-1117 Budapest \\
E-mail: \email{fodor@theorie.physik.uni-wuppertal.de}}
\author{\speaker{Christa Guse}\\
Department of Physics, University of Wuppertal, Gauss 20, D-42119, Germany \\
E-mail: \email{chay-chrissi@web.de}}
\author{Sandor D. Katz\\
Institute for Theoretical Physics, E$\ddot{\mbox{o}}$tv$\ddot{\mbox{o}}$s University, P$\acute{\mbox{a}}$zm$\acute{\mbox{a}}$ny 1, H-1117 Budapest \\
E-mail: \email{katz@bodri.elte.hu}}
\author{Kalman K. Szabo\\
Department of Physics, University of Wuppertal, Gauss 20, D-42119, Germany \\
E-mail: \email{szaboka@general.elte.hu}}

\abstract{
The grand canonical potential for a static quark-antiquark pair at non-vanishing chemical potential was determined at several temperatures. The dynamical staggered simulations were carried out with 2+1 physical quarks along the Lines of Constant Physiscs (LCP).
}

\FullConference{XXIIIrd International Symposium on Lattice Field Theory\\
25-30 July 2005\\
Trinity College, Dublin, Ireland}

\begin{document}

\section{Introduction}
Forces among infinitely heavy quarks separated by distance $r$ are of significance both at $T=0$ and when they are surrounded by an interacting medium. This paper is an extension of \citep[]{Fodor:2004ft}, where a reweighting technique was used to compute the grand canonical potential of a $q \bar{q}$ pair surrounded by a medium of finite chemical potential $\mu$ along the transition line. In this work two other temperatures, one above and one below the transition temperature, will be examined. All simulations were carried out for 2+1 flavours with physical quark masses along a line of constant physics (LCP).

\section{Line of constant physics}
As we want to perform all our finite temperature simulations on lattices of size $4 \times 12^3$, we have to control the temperature $T = 1/(4 a)$ by the lattice spacing $a$. But on changing $a$ it is also necessary to tune the mass parameters $\hat{m}_{u,d}$ and $\hat{m}_s$ accordingly, where the hat denotes the quantities in lattice units. To determine this so called line of constant physics, we carried out zero temperature simulations for several values of the gauge coupling $\beta$ and  of the mass parameters (see Table 1). We used $2+1$ flavours with dynamical staggered fermions. A few hundred configurations were generated for each parameter set using the R-algorithm and setting the microcanonical stepsize to half of the light quark mass.
\begin{table}[h!]
\label{par}
\center
\begin{tabular}{|c|c|c|}
\hline
$\beta$ & $\hat{m}_{u,d}$ & $\hat{m}_s$ \\ \hline \hline
\small 5.09 & \small 0.020, 0.040, 0.060 & \small 0.250 \\ \hline
\small 5.16 & \small 0.020, 0.040, 0.060 & \small 0.250 \\ \hline
\small 5.19 & \small 0.020, 0.040, 0.060 & \small 0.250 \\ \hline
\small 5.32 & \small 0.010, 0.020, 0.030 & \small 0.168 \\ \hline
\small 5.42 & \small 0.010, 0.020, 0.030 & \small 0.126 \\ \hline
\small 5.50 & \small 0.008, 0.016, 0.024 & \small 0.100 \\ \hline
\small 5.57 & \small 0.008, 0.016, 0.024 & \small 0.100 \\ \hline
\end{tabular}
\caption{Parameters used for the zero temperature simulations.}
\end{table}
We measured $\hat{m_\pi}$, $\hat{m_\rho}$, $\hat{m_K}$, $\hat{r_0}$ and $\hat{\sigma}$ at 3 quark masses for each $\beta$ in order to extrapolate to physical masses. The lattice sizes ranged from $24 \times 12^3$ for coarse lattices over $32 \times 16^3$ to $48 \times 24^3$ for finer lattices, so that even for the lightest quark masses $m_\pi \, L > 4$, where $L$ is the spatial extent of the lattice. The ratio $m_\pi/m_\rho$ was set to its physical value ($0.179$) in order to fix $\hat{m}_{u,d}$ as a function of $\beta$ and $m_\pi/m_K$ was set to its physical value ($0.278$) to determine $\hat{m}_{u,d}/\hat{m}_s$. The values of $\hat m_{ud}$ and $\hat m_s$ as the function of $\beta$ are shown in Figure \ref{LCP1}. A weighted average of $\hat{m}_\rho$, $\hat{r}_0$ and $\hat{\sigma}$ was used to set the scale (Figure \ref{LCP2}). We extrapolated the ratio of the strange and light quark masses to the continuum limit (assuming $O(a^2)$ scaling) and got
\begin{equation}
\frac{m_s}{m_{u,d}} = 24.2(7) \quad \mbox{,}
\end{equation}
which is in agreement with chiral perturbation theory \citep{Gasser:1982ap}.
\begin{figure}[h!]
\center
\begin{minipage}[t]{0.49\textwidth}
\input{./LCP2.tex}
\caption[]{$20 \,\,\hat{m}_{u,d}$ and $\hat{m}_s$ as a function of $\beta$ so \\ that the $m_\pi/m_\rho$ and $m_\pi/m_K$ are kept constant. \\ The black points denote the parameters used \\ in the finite temperature simulations.}
\label{LCP1}
\end{minipage}
\begin{minipage}[t]{0.5\textwidth}
\input{./Scale.tex}
\caption[]{The scale $1/a$ as a function of $\beta$. \\ The black points denote the parameters used \\ in the finite temperature simulations.}
\label{LCP2}
\end{minipage}
\end{figure}
A similar line of constant physics using 3 flavours is presented in \citep{Aoki:2005vt}.

\section{Reweighting technique}
\begin{figure}[h!]
\center
\input{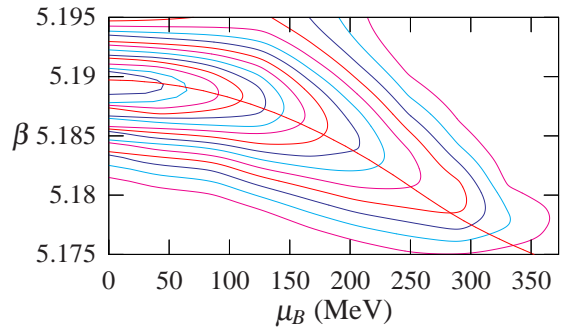}
\caption[]{Overlap in the $\mu$-$\beta$ plane around the transition line: The transition line (taken from \citep{Fodor:2004nz}) matches the line of maximal overlap quite well, as expected \citep{Ejiri:2004yw}.}
\label{sign}
\end{figure}

As the fermion determinant is no longer positive for non-zero chemical potential any kind of importance sampling is hindered. We used the overlap-improving multiparameter reweighting technique \citep{Fodor:2001au}. The expectation value of an operator $\mathcal{O}$ at non-vanishing chemical potential can be rewritten as the ratio of two expectation values at $\mu_0=0$, where importance sampling can be used to generate the configurations.
\begin{eqnarray}
\label{reweighting}
\left< \mathcal{O} \right>_{\mu, \beta} = \frac{\int d\mathcal{U}\,\mathcal{O}\,\omega}{\int d\mathcal{U}\, \omega} = \frac{\int d\mathcal{U}\,\mathcal{O}\,\omega'\,[\det M(\mu_0)]^{n_f/4}\, e^{-S_g(\beta_0)}}{\int d\mathcal{U}\, \omega'\,[\det M(\mu_0)]^{n_f/4}\, e^{-S_g(\beta_0)}} = \frac{\left< \mathcal{O}\,\omega' \right>_{\mu_0, \beta_0}}{\left< \omega' \right>_{\mu_0, \beta_0}} \,\mbox{,}
\end{eqnarray}
\begin{eqnarray}
\mbox{where}\qquad \omega = [\det M(\mu)]^{n_f/4}\, e^{-S_g(\beta)} \,\mbox{,} \qquad \omega' = \frac{[\det M(\mu)]^{n_f/4}\, e^{-S_g(\beta)}}{[\det M(\mu_0)]^{n_f/4}\, e^{-S_g(\beta_0)}} \,\mbox{.} \nonumber
\end{eqnarray}
The details of the determinant evaluation can be found in \citep{Fodor:2001pe}.
As shown by eq. (\ref{reweighting}) a reweighting in other parameters like $\beta$ is also possible and it has the advantage that it can maximize the overlap or trace out the phase transition line. There are however two limitations to the reweighting approach. First the overlap between the sample generated at $\mu=0$ and the target sample at $\mu\neq0$ decreases with growing $\mu$.
One can define a quantity which describes the overlap in several ways \citep{Csikor:2004ik}. In this work it is defined as the ratio of the number of configurations in two different samples $\#(set')/\#(set)$. Here (set) marks the sample generated at $\mu=0$ and (set') the sample obtained after an accept-reject step with probability $\min (1, |\omega'_i/\omega'_{i-1}|)$. The index $i$ stands for the $i$th configuration in (set). This overlap can then be maximized by searching for the optimal path in the $\mu$-$\beta$ plane, as shown in Figure \ref{sign}. The line of maximal overlap for $T$ around $T_C$ matches the transition line, whereas for $T=0.9\,T_C$ and $T=1.33\,T_C$ it is approximately a line of constant temperature at least in the area where the reweighting is reliable.
A second limitation, which is not measured by the overlap defined above is the sign problem. Its impact on the measurements is contained in the jackknife error.

\section{Static $q\bar{q}$ free energy at $T\neq0$ and $\mu\neq0$}
We used the Polyakov-loop correlator to compute $\phi_{q\bar{q}}(T,\mu,r)$, which denotes the difference in the free energies with and without a static $q\bar{q}$ pair separated by distance $r$.
\begin{eqnarray}
e^{-\phi_{q\bar{q}}(T,\mu,r)/T+C(a)} = \left< L(r) L^\dag(0) \right>_{T,\mu} \quad \mbox{where} \qquad L(r) = \frac{1}{3} \mbox{tr} \prod_{\tau=0}^{N_\tau=1} \mathcal{U}_4(\tau,r)
\end{eqnarray}
Here $C$ is a renormalization constant, which was fixed by first normalizing the zero temperature potential demanding $V_{q\bar{q}}(r_0) = 0$ and then removing this shift from the free energy at $T\neq0$ and $\mu\neq0$, as it is the self-energy of the static quark pair at a given lattice spacing. Note that this type of renormalization is manifestly temperature independent.
The finite temperature simulations were carried out using dynamical staggered quarks with $n_f=2+1$ flavours on $4 \times 12^3$ lattices. The R-algorithm was used, the microcanonical stepsize was half of the light quark mass. We generated about 150000 configurations and calculated the determinant after every 50 trajectories giving us about 1500 independent configurations at $T=0.9 T_c$ and $T=1.33 T_c$ and about 3000 at $T=T_c$.
Figure \ref{POT1}-\ref{POT3} show $\phi_{q\bar{q}}$ for the three temperatures. The potential at $\mu_B = 0$ MeV is plotted in black (upper curves), that at $\mu_B = 270$ MeV in red (lower curves). For all temperatures only slight changes can be observed between the potential at zero and finite $\mu$. This is expected as the reweighting was performed along the lines of maximal overlap. For all temperatures the potential reaches an asymptotic value at large distances, due to the screening of the medium. Generally it can be stated, that for higher $\mu$ values the potential reaches a lower asymptotic value. Looking at the different scales of the axes one observes that the change in the potential increases with falling temperature, although being generally rather small. It can be observed that in the deconfined phase the potential flattens out at shorter distances. This is also demonstrated in Figure \ref{POT2}, where in blue (lowest curve) the potential at $\mu_B = 270$ MeV reweighted not along the transition line but along the line of constant temperature is drawn. Here the reweighting line cuts into the deconfined phase. In Figure \ref{POT4} the free energy at a fixed distance is plotted as a function of the baryonic chemical potential. Particulary in the confined phase and also along the transition line one observes an increase of the errorbars due to the sign problem. A similar study using a Taylor expansion in $(\mu/T)$ with 2 degenerate quark flavours and masses corresponding to $m_\pi \approx 770$ MeV \citep{Doring:2005ih} was also presented at the symposium. A thorough comparison is difficult, but we observe that the shifts $[\phi_{q\bar{q}}(T,0,\infty)-\phi_{q\bar{q}}(T,\mu,\infty)]$ are in general agreement, although our values tend to be about $15 \%$ smaller.

\section{Conclusions}
{We have calculated the grand canonical potential of a static $q\bar{q}$ pair for three different lines on the $T$-$\mu$ plane: along the transition line as well as along one line above and along one line below it. All computations were done at physical quark masses on the line of constant physics using an overlap improving multiparameter reweighting in the gauge coupling $\beta$ and in the chemical potential $\mu$. We observed only a small change in the potential with growing $\mu$ in the parameter range where our analysis was carried out.}

{\small{\bf Acknowledgments} The computations were carried out at E$\ddot{o}$tv$\ddot{o}$s University on the 330 processor PC cluster of the Institute for Theoretical Physics and the 1024 processor PC cluster of Wuppertal University, using a modified version of the publicly available MILC code \citep{MILC} and a next-neighbour communication architecture \citep{Cluster}.}

\newpage

\begin{figure}[h!]
\center
\begin{minipage}[t]{0.49\textwidth}
\input{./potential4.tex}
\caption[]{\small The grand canonical potential for \\ $T=0.9\,\,T_C$: The upper curve (black) denotes \\ the potential at $\mu_B = 0$ MeV, whereas the \\ lower curve (red) is the potential reweighted \\ along the best reweighting line to $\mu_B = 270$ MeV. \\ In order to reduce the noise and to make the plot \\ more readable, a box averaging was used. The \\ box size ranged from 0.008 fm, at short distances \\ to 0.2 fm at large distances.}
\label{POT1}
\end{minipage}
\begin{minipage}[t]{0.5\textwidth}
\input{./potential1.tex}
\caption[]{\small The grand canonical potential for $T=1.0\,\,T_C$: In this case the upper curve (black) denoting the potential at $\mu_B = 0$ MeV and the slightly lower curve (red) denoting the potential reweighted along the best reweighting line lie nearly on top of each other, due to the reweighting along the best reweighting line. The lower curve (blue) is the potential reweighted along a line of constant temperature to $\mu_B = 270$ MeV.}
\label{POT2}
\end{minipage}
\end{figure}

\begin{figure}[h!]
\center
\begin{minipage}[t]{0.49\textwidth}
\input{./potential2.tex}
\caption[]{\small The grand canonical potential \\ for $T=1.33\,\,T_C$: The upper curve (black) denotes \\ the potential at $\mu_B = 0$ MeV, whereas the \\ lower curve (red) is the potential reweighted \\ along the best reweighting line to $\mu_B = 270$ MeV.}
\label{POT3}
\end{minipage}
\begin{minipage}[t]{0.5\textwidth}
\input{./potential3.tex}
\caption[]{\small $\phi_{q\bar{q}}$ as a function of $\mu_B$ for a fixed \\ distance $\hat{r}=\sqrt{10}$.}
\label{POT4}
\end{minipage}
\end{figure}

\end{document}